\documentclass[letterpaper]{article} 
\usepackage[draft]{aaai2026}  
\usepackage{times}  
\usepackage{helvet}  
\usepackage{courier}  
\usepackage[hyphens]{url}  
\usepackage{graphicx} 
\def\UrlFont{\rm}  
\usepackage{natbib}  
\usepackage{caption} 
\usepackage{algorithm}
\usepackage{algorithmic}
\usepackage{amsmath}
\usepackage{multirow}
\usepackage{booktabs}
\usepackage{amsfonts}

\usepackage{newfloat}
\usepackage{listings}
\DeclareCaptionStyle{ruled}{labelfont=normalfont,labelsep=colon,strut=off} 
\floatstyle{ruled}
\newfloat{listing}{tb}{lst}{}
\floatname{listing}{Listing}
\title{Say More with Less: Variable-Frame-Rate Speech Tokenization via Adaptive Clustering and Implicit Duration Coding}
\author{
    Rui-Chen Zheng, Wenrui Liu, Hui-Peng Du, Qinglin Zhang\textsuperscript{\rm 1}, Chong Deng\textsuperscript{\rm 1}, Qian Chen\textsuperscript{\rm 1}, Wen Wang\textsuperscript{\rm 1}, Yang Ai\thanks{Corresponding Author.}, Zhen-Hua Ling
}
\affiliations{
    \textsuperscript{\rm 1} Tongyi Fun Team, Alibaba Group\\
    zhengruichen@mail.ustc.edu.cn
}

\usepackage{bibentry}

\begin{document}

\maketitle

\begin{abstract}
Existing speech tokenizers typically assign a fixed number of tokens per second, regardless of the varying information density or temporal fluctuations in the speech signal. This uniform token allocation mismatches the intrinsic structure of speech, where information is distributed unevenly over time. 
To address this, we propose \textbf{VARSTok}, a \textbf{VA}riable-frame-\textbf{R}ate \textbf{S}peech \textbf{Tok}enizer that adapts token allocation based on local feature similarity. 
VARSTok introduces two key innovations: (1) a temporal-aware density peak clustering algorithm that adaptively segments speech into variable-length units, and (2) a novel implicit duration coding scheme that embeds both content and temporal span into a single token index, eliminating the need for auxiliary duration predictors. 
Extensive experiments show that VARSTok significantly outperforms strong fixed-rate baselines. Notably, it achieves superior reconstruction naturalness while using up to 23\% fewer tokens than a 40 Hz fixed-frame-rate baseline. VARSTok further yields lower word error rates and improved naturalness in zero-shot text-to-speech synthesis. To the best of our knowledge, this is the first work to demonstrate that a fully dynamic, variable-frame-rate acoustic speech tokenizer can be seamlessly integrated into 
downstream speech language models.
\end{abstract}

\begin{links}
    \link{Project page with samples and code}{https://zhengrachel.github.io/VARSTok}
    \link{Extended version with appendix}{https://arxiv.org/abs/2509.04685}
\end{links}

\section{Introduction}
Speech tokenization has become central to modern speech modeling, powering advances in neural audio codecs \cite{zeghidour2021soundstream, wu2023audiodec, ai2024apcodec, huang2024repcodec}, generative speech synthesis \cite{zhang2024speechtokenizer, chen2025neural}, and multimodal large language models (LLMs) \cite{du2024cosyvoice, chen2025minmo}. By converting continuous speech signals into discrete token sequences, speech tokenization bridges the gap between raw waveform and token-based language modeling architectures, thus enabling the application of LLMs to speech data.

Existing speech tokenization methods generally fall into three types: semantic tokenizers \cite{baevski2020wav2vec, hsu2021hubert} that capture high-level linguistic content; acoustic tokenizers \cite{kumar2023high, xin2024bigcodec, liu2025analyzing} that preserve fine-grained signal fidelity; and hybrid approaches \cite{zhang2024speechtokenizer, ye2025codec} that combine the merits of both. While recent acoustic tokenizers have shown impressive results, they typically operate at fixed frame rates (e.g., 75 Hz), uniformly allocating tokens across time, ignoring the underlying content or information density.

However, natural speech is inherently non-uniform over time \cite{keshishian2021understanding}. Segments with silence or stable vowels are acoustically redundant, while others with rapid articulatory transitions or expressive prosody carry dense information \cite{van2017information, dieleman2021variable}. 
Fixed-rate tokenizers fails to adapt to this variability, leading to inefficient token usage and poor alignment with temporal dynamics. This misalignment not only results in suboptimal compression but also hinders the ability of downsteam speech language models (LMs) to learn natural prosody and rhythm. Addressing this challenge requires a paradigm shift from fixed-rate to content-aware, dynamic tokenization. 

To this end, we propose VARSTok,  a fully dynamic, variable-frame-rate speech tokenizer that adaptively allocates tokens based on local feature similarity. It introduces a temporal-aware clustering algorithm to segment speech into variable-length units. To support downstream language modeling  without auxiliary duration predictors, we design an implicit duration coding scheme that embeds both content and duration into a single token index. 
In contrast to prior work \cite{zhang2025unlocking}, VARSTok enables fully dynamic token allocation without the need for hierarchical fusion or predefined temporal resolutions. Crucially, the resulting duration-aware tokens can be seamlessly used in autoregressive speech LMs without modification.


We validate the effectiveness of VARSTok across speech reconstruction, semantic evaluation, and text-to-speech (TTS) language modeling. 
Experimental results show that despite operating at a lower average token rate of 30.95 Hz, VARSTok achieves better reconstruction quality and semantically richer representations than a 40 Hz fixed-rate baseline. In TTS, it achieves improved naturalness while maintaining comparable or even lower word error rates (WER). 
To the best of our knowledge, this is the first work to demonstrate that a fully dynamic, variable-frame-rate acoustic speech tokenizer can be seamlessly integrated into downstream speech LMs without requiring architectural modifications.

In summary, our contributions are:
\begin{enumerate}
    \item We introduce VARSTok, the first fully dynamic, variable-frame-rate speech tokenizer that can be seamlessly integrated into downstream autoregressive speech LMs. 
    \item 
    We propose a temporal-aware clustering algorithm and a novel implicit duration coding scheme that jointly produce efficient, duration-aware token sequences, eliminating the need for auxiliary duration predictors or hierarchical token streams.
    \item We demonstrate that VARSTok achieves competitive performance across three tasks, consistently outperforming strong fixed-rate baselines while using fewer tokens.
\end{enumerate}

\section{Related Work}

\subsection{Speech Tokenization}
Speech tokenization has gained increasing attention for its role in neural audio compression, generative speech modeling, and multimodal LLMs \cite{cui2024recent}. Existing methods broadly fall into three categories.

\textbf{Semantic tokenizers} extract high-level linguistic units by applying clustering or vector quantization (VQ) \cite{van2017neural} over features learned from large-scale self-supervised pretraining \cite{hsu2021hubert, baevski2020wav2vec, chen2022wavlm}. While effective for speech understanding tasks \cite{yoon2024hubert, sharma2022multi, fang2024llama}, they discard fine-grained acoustic details and do not support direct waveform reconstruction, limiting their utility in generative applications such as speech synthesis or editing.

\textbf{Acoustic tokenizers} focus on preserving waveform fidelity by mapping raw speech to discrete tokens using residual VQ \cite{zeghidour2021soundstream,defossezhigh,kumar2023high} Compared to semantic tokens, acoustic tokens retain richer speech details and can be directly used in generation pipelines without a separate token-to-waveform model. 
Recent approaches have focused on single-codebook designs \cite{xin2024bigcodec, zhai2025one, zengscaling} to improve token efficiency and compatibility with downstream speech LMs \cite{borsos2023audiolm, borsos2023soundstorm}. WavTokenizer \cite{ji2024wavtokenizer} achieves strong reconstruction quality at fixed frame rates of 40 Hz or 75 Hz using a single codebook. However, they still rely on fixed-rate token assignment, failing to account for the temporal variability of speech signals.

\textbf{Hybrid tokenizers} aim to combine the merits of both by distilling linguistic content into acoustic embeddings \cite{zhang2024speechtokenizer, ye2025codec, ye2025llasa}. While enabling high-quality synthesis with linguistic control, they often require multiple codebooks, hierarchical token streams, or auxiliary modules, increasing system complexity and reducing generality.

VARSTok builds upon the acoustic tokenizer paradigm but departs from the fixed-rate framework by operating at a variable frame rate. Instead of assigning tokens uniformly in time, it adaptively segments speech based on local feature similarity, producing a dynamic token stream that better reflects the temporal variation of speech. This enables improved token efficiency and greater alignment with speech structure compared to conventional fixed-rate approaches.

\subsection{Adaptive Compression in Speech}
Recent work has explored dynamic speech compression primarily in the context of semantic representation learning. For example, SD-HuBERT \cite{cho2024sd} uses sentence-level self-distillation to induce syllabic organization. Other approaches employ learnable temporal pooling \cite{dieleman2021variable}, reinforcement learning \cite{cuervo2022variable}, or syllable-aligned self-supervised representations \cite{baadesyllablelm, chosylber} for semantic boundary discovery. However, these methods are designed for semantic abstraction and discard fine-grained acoustic information, resulting in non-reconstructable representations that are inappropriate for waveform generation, transmission, or any task requiring high-fidelity reconstruction.

High-fidelity acoustic tokenizers predominantly, on the other hand, largely adopt a fixed-frame-rate paradigm, leaving dynamic alternatives underexplored. The most related work to ours is TFC \cite{zhang2025unlocking}, which introduces variable temporal resolution into neural speech codec built upon multi-codebook residual VQ. TFC adaptively select among three predefined frame rates (75 Hz, 37.5 Hz, 18.75 Hz) based on entropy-based information density estimation, enabling flexible bitrate control and improved reconstruction under constrained token budgets. 
However, TFC relies on a fixed set of temporal granularities and constructs final representations by hierarchically fusing features across coarse-to-fine scales. Although the resulting frame rate varies, each token still adheres to one of the predefined resolutions, making the system pseudo-dynamic. Moreover, TFC ignores the modeling of token duration, limiting its utility in autoregressive modeling or alignment-sensitive tasks. 

In contrast, VARSTok achieves fully dynamic token allocation through temporal-aware clustering without relying on fixed downsampling schedules. The proposed implicit duration coding further enables its use for downstream speech LMs without requiring auxiliary duration predictors. 
While both TFC and VARSTok aim to develop variable-frame-rate speech tokenization for high-fidelity reconstruction, our method offers a more flexible and complete solution, better aligned with the needs of downstream speech LMs. 

\begin{figure*}[t]
    \centering
    \includegraphics[width=0.85\linewidth]{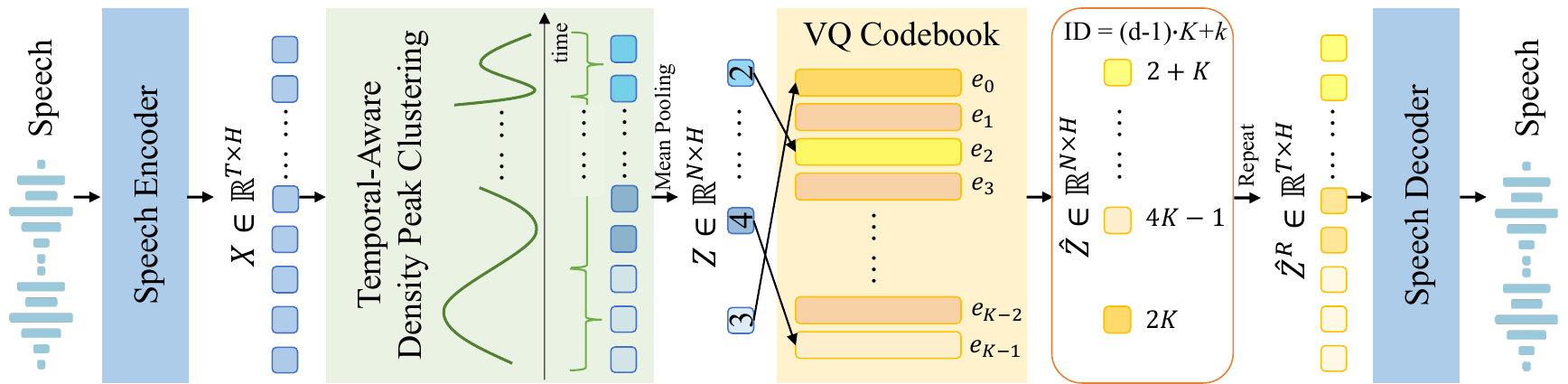}
    \caption{
        \textbf{Overview of VARSTok.}
        Input waveform is converted into frame-level embeddings via a speech encoder.
        Temporal-aware density peak clustering adaptively segments them into variable-length clusters based on similarity and temporal continuity.
        Each cluster is mean-pooled and quantized using a VQ codebook to produce a discrete token whose index encodes both content and duration (i.e., number of frames spanned).
        Each token embedding is expanded back to frame-level representations according to its duration and passed to the decoder for waveform reconstruction.
    }
    \label{fig:overview}
\end{figure*}

\section{Proposed Methods}

\subsection{Overview}
We propose VARSTok, a fully dynamic, variable-frame-rate speech tokenizer that adaptively allocates tokens based on local speech feature similarity. Unlike conventional fixed-rate tokenizers, VARSTok reduces token usage in redundant regions while assigning more tokens to segments with rich variation, leading to more efficient modeling and better alignment with the intrinsic fluctuations of speech.

As illustrated in Figure \ref{fig:overview}, VARSTok comprises four major components: a speech encoder, a temporal-aware density peak clustering module, a VQ module, and a speech decoder.
First, the speech encoder transforms raw waveforms into frame-level embeddings $\mathbf{X} =[\mathbf{x}_1, \dots, \mathbf{x}_T] \in \mathbb{R}^{T\times H}$, where $T$ is the number of frames determined by the encoder's base frame rate, and $H$ is the embedding dimension. 
Adjacent frames are then adaptively grouped into $N$ variable-length clusters $\mathcal{C} = \{\mathcal{C}_1,\dots,\mathcal{C}_N\}$ based on local feature similarity, using the temporal-aware density peak clustering module. Each cluster $\mathcal{C}_n$ is summarized into a mean-pooled embedding $\mathbf{z}_n \in \mathbb{R}^H$and quantized using a VQ module with a single codebook $\mathcal{E}=\{\mathbf{e}_0,\dots,\mathbf{e}_{K-1}\}$ of size $K$, where $\mathbf{e}_k$ denotes the $k$-th codebook entry. 
The quantized embedding $\hat{\mathbf{z}}_n$ for each cluster is paired with its duration $d_n$, representing the number of original frames spaned by the cluster. 
This duration is further integrated into the token index by the proposed implicit duration coding scheme, allowing each token to jointly encode both content and temporal span, thereby preserving alignment without requiring auxiliary duration predictors. 
Before decoded by a speech decoder, the quantized embeddings $\hat{\mathbf{Z}}=[\hat{\mathbf{z}}_1 ,\dots, \hat{\mathbf{z}}_N]$ are expanded according to their durations $\mathbf{d}=[d_1, \dots, d_N]$ to form a frame-level sequence $\hat{\mathbf{Z}}^R = [\hat{\mathbf{z}}^R_1 ,\dots, \hat{\mathbf{z}}^R_T]\in\mathbb{R}^{T\times H}$, restoring the temporal structure for high-fidelity waveform reconstruction. 

The overall design enables VARSTok to produce compact and temporally-aligned token sequences that can be directly used in downstream speech LMs, achieving high reconstruction quality with significantly improved token efficiency. We describe each component in detail in the following sections.

\subsection{Model Architecture}
VARSTok adopts an encoder-VQ-decoder architecture following the design of WavTokenizer \cite{ji2024wavtokenizer}, which achieves advanced performance with a single codebook.

\subsubsection{Speech Encoder and Decoder}

The speech encoder begins with a 1D convolutional layer followed by four convolutional blocks, each containing a residual unit and a strided convolution for temporal downsampling while doubling the number of channels at each stage. The downsampled features are further processed by a two-layer bidirectional LSTM and a projection layer, producing frame-level embeddings $\mathbf{X}$. ELU activations are used throughout. 
The speech decoder adopts an improved architecture combining attention layers, ConvNextV2 \cite{woo2023convnext} blocks, and an inverse Fourier transform-based upsampling module. 
In VARSTok, the decoder receives an expanded sequence of quantized cluster embeddings $\hat{\mathbf{Z}}^R $ obtained by repeating each quantized cluster embedding  $\hat{\mathbf{z}}_n $ according to their durations $d_n$, and finally reconstructs the original waveform.

\subsubsection{VQ Module}

We adopt the VQ module from WavTokenizer, which employs a single codebook with $K=4096$ entries. Each cluster embedding $\mathbf{z}_n\in\mathbb{R}^H$ is mapped to its nearest codebook entry $\mathbf{e_{k_n}} \in \mathcal{E}$ based on L2 distance after factorization. The codebook is updated using exponential moving averages to encourage stability and high codebook usage. To further improve token diversity and prevent codebook collapse, random awakening is applied during training.

\subsection{Temporal-Aware Density Peak Clustering}

\begin{algorithm}[t]
\caption{Temporal-Aware Density Peak Clustering}
\label{alg:tadpc}
\begin{algorithmic}[1]
\REQUIRE Embeddings $\mathbf{X} \in \mathbb{R}^{T \times D}$, neighbors $m$, threshold $\tau$, penalty $\beta$, max span $S_{\max}$
\ENSURE Cluster embeddings $\mathbf{Z} = \{\mathbf{z}_1, \dots, \mathbf{z}_N\}$ and durations $\mathbf{d}=\{d_1, \dots, d_N\}$

\STATE Compute similarity $\phi(\mathbf{x}_i, \mathbf{x}_j)$ for all pairs
\STATE Compute local density $\rho_i$ and peak distance $\delta_i$
\STATE Compute peak score $s_i = \rho_i \cdot \delta_i$
\STATE Initialize \texttt{assigned} = \texttt{False} for all frames

\WHILE{some frames are unassigned}
    \STATE Select seed $i^* = \arg\max s_i$ among unassigned
    \STATE Initialize cluster $\mathcal{C}_n = \{i^*\}$

    \FOR{offset $t = \pm 1$ to $\pm S_{\max}$}
        \STATE Check similarity and score: $\phi(\mathbf{x}_{i^*}, \mathbf{x}_t) - \beta\cdot s_t > \tau$
        \STATE If valid and unassigned, add $t$ to $\mathcal{C}_n$, else \texttt{break}
    \ENDFOR

    \STATE Sort $\mathcal{C}_n$; compute mean $\mathbf{z}_n = \frac{1}{|\mathcal{C}_n|} \sum_{t \in \mathcal{C}_n} \mathbf{x}_t$, duration $d_n = |\mathcal{C}_n|$
\ENDWHILE
\RETURN $\mathbf{Z} = [\mathbf{z}_1, \dots, \mathbf{z}_N]$, $\mathbf{d}=[d_1, \dots, d_N]$
\end{algorithmic}
\end{algorithm}

To adaptively segment the encoder output $\mathbf{X}$ into variable-length units, we propose a temporal-aware density peak clustering algorithm. 
Unlike prior clustering-based methods \cite{wu2024towards} designed for unordered inputs like image patches, speech features are inherently sequential and clustering must respect temporal continuity. 
Our proposed algorithm enforces strict time-order constraints to produce segments that are both semantically meaningful and temporally aligned. The full algorithm is detailed in Algorithm \ref{alg:tadpc}.

We first identify potential cluster centers among all the frames by calculating the local density $\rho_i$ and peak distance $\delta_i$ of each frame.
The local density $\rho_i$ measures how closely a frame is surrounded by similar neighbors in the embedding space:
\begin{equation}
    \rho_i = \exp{(\frac{1}{m}\sum_{j\in \text{KNN}(i)}\phi(\mathbf{x}_i, \mathbf{x}_j))},
\end{equation}
where $\text{KNN}(i)$ are the indices of the $m$-most-similar frames of $x_i$ and $\phi(\cdot, \cdot)$ denotes the normalized cosine similarity bween two frames:
\begin{equation}
    \phi(\mathbf{x}_i, \mathbf{x}_j) = \frac{1+<\mathbf{x}_i, \mathbf{x}_j>}{2}.
\end{equation}
Intuitively, a higher $\rho_i$ indicates $\mathbf{x}_i$ is in a dense region of the embedding space.
The peak distance $\delta_i$ measures a frame's separation from regions of higher density:
\begin{equation}
    \delta_i =
\begin{cases}
\min\limits_{j: \rho_j > \rho_i} \, 1-\phi(\mathbf{x}_i, \mathbf{x}_j), & \text{if such } j \text{ exists}, \\
\max\limits_{j} \, 1-\phi(\mathbf{x}_i, \mathbf{x}_j), & \text{otherwise}.
\end{cases}
\end{equation}
A larger $\delta_i$ implies that $\mathbf{x}_i$ is an isolated peak in the density landscape.
The final peak score is defined as:
\begin{equation}
    s_i = \rho_i \cdot \delta_i .
\end{equation}
Frames with high $s_i$ are both locally dense and relatively isolated, making them ideal candidates to seed new clusters.

With peak scores $\{s_i\}_{i=1}^{T}$ computed, we form clusters greedily. In each step, we select the unassigned frame $i*$ with the highest peak score $s_{i*}$ to initialize a new cluster $\mathcal{C}_n=\{i*\}$. The cluster is then expanded bidirectionally from the seed $i*$.
A candidate frame $t$ is added to the current cluster $\mathcal{C}_n$ only if it meets two conditions. 
First it must satisfy a similarity criterion:
\begin{equation}
    \phi(\mathbf{x}_{i^*}, \mathbf{x}_t) - \beta \cdot s_t > \tau,
\end{equation}
where $\beta$ is a penalty factor and $\tau$ is a manually defined similarity threshold. The term $-\beta\cdot s_t$ penalizes adding frames that are themselves strong cluster seeds.
Second and crucially for speech, the process is temporal-aware: $t$-th frame considered for inclusion only if its immediate temporal neighbor (i.e., $t-1$ for forward expansion and $t+1$ for backward expansion) has already been assigned to the current cluster $\mathcal{C}_n$. 
This constraint ensures that all clusters are temporally contiguous segments. The expansion stops if either condition fails or if the cluster span reaches a predefined maximum $S_{\max}$.
Once the expansion for cluster $\mathcal{C}_n$ is complete, we compute its mean-pooled embedding $\mathbf{z}_n=\frac{1}{|\mathcal{C}_n|} \sum_{t \in \mathcal{C}_n}\mathbf{x} _t$ and record its span $d_n=|\mathcal{C}_n|$. 
This process repeats until all frames are assigned, yielding a variable-length sequence of cluster embeddings $[\mathbf{z}_1,\dots, \mathbf{z}_N]$.

\subsection{Implicit Duration Coding via Extended Index}
A key challenge in variable-frame-rate speech tokenization is managing the duration of each token. Accurate duration modeling is critical for faithfully reconstructing the temporal structure of speech and ensuring proper alignment with other modalities in downstream speech LMs. 
A natural solution is to incorporate a separate FastSpeech-style duration predictor \cite{ren2019fastspeech, renfastspeech}. However, this increases architectural complexity and breaks the simplicity of an end-to-end tokenizer. More importantly, we found emprically that jointly training a learned duration predictor within the tokenizer pipeline leads to severe optimization instability and poor convergence. To address this, we propose a simple yet effective implicit duration coding scheme that embeds both content identity and temporal span into a single token index. This design preserves temporal alignment without requiring auxiliary predictors, while maintaining the stability and modularity of the overall tokenizer architecture.

The core idea is to expand the VQ codebook index space by a factor of $S_{\max}$, the maximum allowable cluster duration, resulting in a conceptual vocabulary of $K \cdot S_{\max}$ unique token IDs. In practice, only the original single codebook $\mathcal{E}$ of size $K$ is instantiated and trained. Specifically, a quantized cluster embedding $\hat{\mathbf{z}}_n$ with VQ codebook index $k_n \in \{0, \dots, K-1\}$ and duration $d_n \in  \{1, \dots, S_{\max}\}$ is mapped to a single, unified token ID:
\begin{equation}
    \text{ID}_n = (d_n-1) \cdot K + k_n .
\end{equation}
During the decoding stage, this process is trivially reversed to recover both the content and duration from the token ID. The duration is recovered via integer division: 
\begin{equation}
    d_n = \left\lfloor \frac{\text{ID}_n}{K} \right\rfloor + 1,
\end{equation}
and the original VQ index is recovered via 
\begin{equation}
    k_n = \text{ID}_n \bmod K.
\end{equation}
The corresponding quantized cluster embeddings $\hat{\mathbf{z}}_n=\mathbf{e}_{k_n}$ is then repeated $d_n$ times to form the input for the speech decoder, restoring the correct temporal resolution for waveform reconstruction.



When applied to downstream speech LMs, the implicit duration coding scheme offers substantial advantages. By embedding both content and duration within a single token, the modeling process is significantly streamlined: the speech LM no longer requires an auxiliary duration predictor to infer temporal spans. The model operates directly on a highly compact sequence of extended token IDs without frame-level repetition or upsampling. This reformulation reduces speech modeling to a standard autoregressive prediction task, where each step selects the next token from an expanded vocabulary of size $K\times S_{\text{max}}$. This design preserves the LM's end-to-end simplicity while remaining lightweight, flexible, and readily scalable with different codebook sizes and maximum-allowed durations.

\begin{table*}[htbp]
  \centering
  \fontsize{9pt}{11pt}\selectfont
  \caption{Speech reconstruction performance of VARSTok compared with single-codebook baselines. Results are reported on the LibriTTS test-clean set. 
  UTMOS, PESQ, and V/UV F1 scores are computed by downsampling the speech to 16 kHz. The STOI for SQCodec is marked as '/' since STOI values for the other systems are calculated at 24 kHz. 
The optimal results below 40Hz frame rate are marked in \textbf{bold}.  }
    \begin{tabular}{c|cc|c|c|cccc}
    \toprule
    Model &$\tau$  &$S_{\max} $  & Frame Rate/Hz$\downarrow$ & Bitrate/kbps$\downarrow$ & UTMOS$\uparrow$ & PESQ$\uparrow$  & STOI$\uparrow$  & V/UV F1$\uparrow$ \\
    \hline
    GT    & /     & /     & /& /     & 4.1185& / & / & /\\ \hline
    DAC   & /     & /     & 100.00& 1.00& 1.4940& 1.2464 & 0.7706 & 0.7941 \\
    WavTokenizer & /     & /     & 75.00& 0.90& 4.0247 & 2.4543 & 0.9188 & 0.9339 \\
    XCodec2.0 & /     & /     & 50.00& 0.80& 3.4727& 1.8659& 0.8635& 0.9136\\ 
    VARSTok & 0.7     & 2     & 46.50& 0.60& 4.0379 & 2.0694 & 0.8935 & 0.9209\\
    SQCodec & /     & /     & 44.44 & 0.75 & 3.9601 & 1.8898 & /& 0.9197 \\ \hline
    BigCodec & /     & /      & 40.00& 0.52 & 3.9802& 1.8796& 0.8653& 0.9133 \\
    WavTokenizer & /     & /     & 40.00& 0.48 & 3.6107 & 1.7075 & 0.8652 & 0.9095 \\
    VARSTok & 0.8     & 4     & 36.81 & 0.52 & \textbf{4.0000} & \textbf{1.8887} & \textbf{0.8814} & \textbf{0.9186} \\
    VARSTok & 0.7     & 4     & 30.95 & 0.43 & 3.8949 & 1.7095 & 0.8601 & 0.9047 \\ 
    VARSTok & 0.6     & 4     & 26.29 & 0.37 & 3.8304 & 1.5855 & 0.8411 & 0.8985 \\
    VARSTok & 0.7     & 8     & 22.38 & 0.34 & 3.6466 & 1.4532 & 0.8203 & 0.8860 \\
    \bottomrule
    \end{tabular}%
  \label{tab:reconstruction}%
\end{table*}%

\subsection{Training Objective}
We follow the same training objective as WavTokenizer, jointly optimizing a mel-spectrogram reconstruction loss, a vector quantization loss, an adversarial loss and a feature matching loss as follows:
\begin{equation}
\mathcal{L} = \lambda_{mel} \mathcal{L}_{mel} + \lambda_q \mathcal{L}_q + \lambda_{adv} \mathcal{L}_{adv} + \lambda_{feat} \mathcal{L}_{feat}.
\end{equation}
More details are provided in the Appendix B.

\section{Experiments}

\subsection{Experimental Setup}
\subsubsection{Datasets and Tasks} We trained VARSTok on the 585-hour LibriTTS corpus \cite{zen2019libritts} at a 24 kHz sampling rate for fair comparison with prior work. Our experiments cover three distinct tasks for comprehensive evaluation:
\begin{itemize}
    \item Speech Reconstruction: We evaluated the speech reconstruction performance of VARSTok using the LibriTTS test-clean subset. Additional results on the more challenging test-other subset are provided in the Appendix G to assess its robustness and generalizability.
    \item Semantic Representation Quality: To evaluate the semantic quality of the learned representations, we adopted the speech portion of the ARCH benchmark\footnote{\url{https://github.com/MorenoLaQuatra/ARCH}.} \cite{la2024benchmarking}, including SLURP \cite{bastianelli2020slurp} (intent classification), EMOVO \cite{costantini2014emovo} and RAVDESS \cite{livingstone2018ryerson} (emotion classification), and AudioMNIST \cite{becker2024audiomnist} (spoken digit recognition). Additional details about the ARCH bechmark are provided in the Appendix E.
    \item Downstream TTS Language Modeling: We evaluated VARSTok's compatibility with downstream speech LMs by training an TTS model based on a decoder-only autoregressive architecture using LibriTTS dataset. Training details are provided in the Appendix F. Its zero-shot TTS performance was assessed on the test set from UniCATS \cite{du2024unicats}.
\end{itemize}

\subsubsection{Baselines} 

We compared VARSTok against a set of strong, single-codebook acoustic tokenizers for a fair and comprehensive evaluation. Our primary baselines are the 40 Hz and 75 Hz variants of WavTokenizer, as VARSTok is built upon its architecture. We evaluated both models using the publicly released small-version checkpoints provided by the authors\footnote{\url{https://github.com/jishengpeng/WavTokenizer}.}. This setup allows for a direct and controlled comparison of our variable-rate mechanism against its fixed-rate counterpart across all three evaluation tasks.

To further situate our model's reconstruction performance within the broader field, we benchmarked it against several other representative tokenizers specifically on the speech reconstruction task. 
These include the single-codebook DAC \cite{kumar2023high} with results quoted from the WavTokenizer paper, alongside BigCodec \cite{xin2024bigcodec} and XCodec2.0 \cite{ye2025llasa} which we trained from scratch on 24 kHz LibriTTS corpus. We also included results from SQCodec as optimistic references. It is important to note that SQCodec and XCodec2.0 benefit from either being trained and evaluated on 16kHz speech using substantially larger datasets or leveraging additional semantic features during inference, making their results not directly comparable. For reproducibility, all detailed configurations for baseline models are provided in the Appendix D. We do not include earlier neural codecs such as EnCodec or SoundStream as baselines, as they rely on multi-codebook or hierarchical token streams that are not directly compatible with downstream speech language modeling. For experimental efficiency, we limited semantic evaluation and TTS comparisons to VARSTok and the 40 Hz variant of WavTokenizer.

\begin{table*}[htbp]
  \centering
  \fontsize{9pt}{11pt}\selectfont
  \caption{Classification accuracy (ACC) and macro-averaged F1 score across four datasets in the speech portion of ARCH benchmark. The optimal results are marked in \textbf{bold}.}
    \begin{tabular}{c|c|cc|cc|cc|cc}
    \toprule
    \multirow{1}[3]{*}{Model}& \multirow{1}[3]{*}{Frame Rate/Hz}& \multicolumn{2}{c|}{EMOVO} & \multicolumn{2}{c|}{RAVDESS} & \multicolumn{2}{c|}{AUDIO MNIST} & \multicolumn{2}{c}{SLUPR} \\
\cline{3-10}          &       & ACC   & F1    & ACC   & F1    & ACC   & F1    & ACC   & F1 \\
    \hline
    WavTokenizer & 40.00& 0.2194 & 0.1676 & 0.2847 & 0.2319 & 0.4597 & 0.4509 & 0.0658 & 0.0055 \\
    VARSTok($\tau=0.8$)& 36.81& \textbf{0.2500}& \textbf{0.1900}& 0.2639 & 0.2241 & 0.5958 & 0.5930& \textbf{0.0755}& \textbf{0.0109}\\
    VARSTok($\tau=0.7$)& 30.95& 0.2347 & 0.1763& \textbf{0.2951} & \textbf{0.2508} & 0.6111 & 0.6078 & 0.0746 & \textbf{0.0109} \\
    VARSTok($\tau=0.6$)& 26.29& 0.2364 & 0.1682 & 0.2674 & 0.2348 & \textbf{0.6207}& \textbf{0.6175}& 0.0729 & 0.0098\\
    \bottomrule
    \end{tabular}%
  \label{tab:arch}%
\end{table*}%

\subsubsection{Evaluation Metrics} 
We employed a comprehensive set of standard metrics to evaluate VARSTok's performance across three tasks.
For speech reconstruction, we report UTMOS \cite{saeki2022utmos} (a neural estimator of Mean-Opinion Score), PESQ (perceptual evaluation of speech quality), STOI (short-time objective intelligibility) and V/UV F1 Score (voiced/unvoiced frame classification accuracy) to assess the naturalness, quality, intelligibility, and prosodic consistency of the reconstructed speech, respectively. 
For VARSTok, the reported ``Frame Rate" is the average token rate per second calculated across the entire test set.
To ensure fair comparison, we calculate the bitrate for VARSTok based on the expanded token space as:
\begin{equation}
\text{Bitrate}=\text{Frame Rate} \times \log_2(K\cdot S_{\max}).
\end{equation}
This formulation accounts for the effective token vocabulary size after expansion due to the inclusion of duration. For baseline models the bitrate is calculated as:
\begin{equation}
    \text{Bitrate} = \text{Frame Rate} \times \log_2 (K).
\end{equation}
For semantic evaluation, we use classification accuracy amd macro-averaged F1 score on the ARCH benchmark to quantify the discriminability of learned representations across the three tasks mentioned above. 
For TTS language modeling, we evaluated word error rate (WER) using the Whisper-large-v3 \cite{radford2023robust} model, speaker similarity via cosine similarity of WavLM\footnote{\url{https://huggingface.co/microsoft/wavlm-base-plus-sv}.} \cite{chen2022wavlm} embeddings, and UTMOS to assess perceptual naturalness of the synthesized speech. Subjective evaluations were also conducted. The setup is described in Appendix I. We report mean opinion scores (MOS) for naturalness and similarity MOS (SMOS) to assess perceived speaker similarity relative to the reference.

\subsubsection{Implementation Details} We initialized VARSTok by loading the encoder, decoder, and VQ codebook from the pretrained small-version 75 Hz WavTokenizer. This initialization provided a solid foundation and facilitated fast convergence when adapting to the proposed variable-frame-rate tokenization scheme. 
No additional losses or auxiliary predictors are used beyond those inherited from WavTokenizer. For the clustering algorithm, we set the penalty coefficient to $\beta = 0.2$ and nearest neighbors $m = 5$. To investigate the impact of clustering granularity, we varied the similarity threshold $\tau \in \{0.6, 0.7, 0.8\}$. We set $S_{\max} = 4$ by default. For the speech reconstruction task, we additionally explored different values of $S_{\max} \in \{2, 4, 8\}$ to study its influence on temporal compression.
A detailed description of our training configuration, including optimizer settings and the learning rate schedule, is provided in Appendix C.

\subsection{Speech Reconstruction}
\subsubsection{Main Results} 
The speech reconstruction results in Table \ref{tab:reconstruction} confirm VARSTok's superior efficiency over fixed-rate methods. For instance, VARSTok configured with $\tau=0.7, S_{\max}=4$ achieves a UTMOS score of 3.8949, surpassing the 40Hz WavTokenizer despite a 23\% reduction in token frame rate.  More strikingly, VARSTok configured with $\tau=0.8, S_{\max}=4$ achieves a UTMOS of 4.0000, nearly on par with the 75 Hz WavTokenizer while using less than half the tokens. These highlight VARSTok’s ability to discard redundancy without sacrificing essential acoustic details. Furthermore, among all models operating below 40 Hz in Tabel \ref{tab:reconstruction}, VARSTok consistently delivers leading performance across all metrics, underscoring its competitiveness in the low-frame-rate regime. Representative speech samples are provided in the Appendix A.

\subsubsection{Analysis of Hyperparameters} We conducted an analysis to investigate the impact of clustering hyperparameters on the rate-quality trade-off and summarized the results in Table \ref{tab:reconstruction}.
The similarity threshold $\tau$ controls the selectivity of cluster expansion from seed frames. A higher $\tau$ enforces stricter similarity, resulting in shorter clusters and thus a higher token rate. 
For instance, increasing $\tau$ from 0.6 to 0.8 boosts the UTMOS score from 3.83 to 4.00, but at the cost of an increased frame rate.
The maximum span $S_{\max}$ constrains the maximum duration of each token, thereby limiting the extent of temporal compression. Larger $S_{\max}$ values allow more aggressive clustering and thus lower token rates, but may underfit fine-grained acoustic detail. We observed that increasing $S_{\max}$ from 2 to 8 reduces the average token rate from 46.50 Hz to 22.38 Hz, but causes severe degradation in UTMOS and PESQ.
Overall, we found $\tau = 0.7$ and $S_{\max} = 4$ to strikes the best balance, achieving a token rate of 30.95 Hz with strong performance.
This analysis validates VARSTok's flexibility, offering granular control over the rate-quality spectrum via an intuitive hyperparameters, which is a capability that fixed-rate models lack.

\subsubsection{Visualization of Token Boundaries} 


\begin{figure*}[t]
    \centering
    \includegraphics[width=0.75\linewidth]{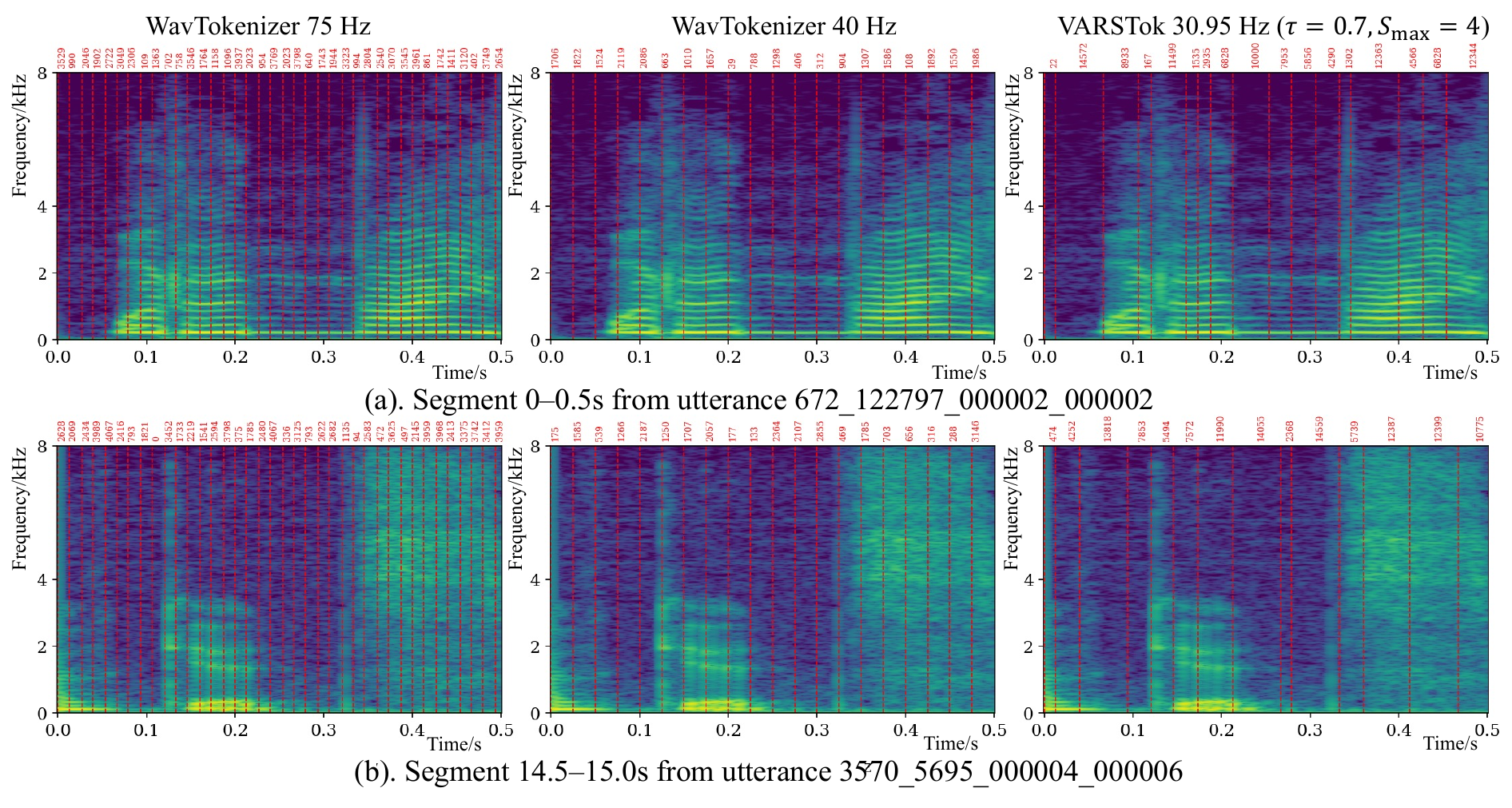}
    \caption{
        \textbf{Token boundary visualization on two speech segments for three tokenizers}
        : WavTokenizer (75 Hz), WavTokenizer (40 Hz), and VARSTok (30.95 Hz). Vertical red lines indicate token boundaries and token IDs are annotated above. 
    }
    \label{fig:visualization}
\end{figure*}

\begin{table*}[tbp]
  \centering
  \caption{Comparison of WER, speaker similarity (SIM), UTMOS between fixed-rate WavTokenizer at a 40Hz frame rate and the proposed VARSTok. The optimal and suboptimal results are marked in \textbf{bold} and \underline{underlined}, respectively.}
    \fontsize{9pt}{11pt}\selectfont
    \begin{tabular}{c|c|ccc|cc}
    \toprule
    Tokenizer & Frame Rate/Hz$\downarrow$ & WER/\%$\downarrow$ & SIM$\uparrow$   & UTMOS$\uparrow$ &MOS$\uparrow$ &SMOS$\uparrow$ \\
    \hline
    WavTokenizer & 40.00& 7.481 & \textbf{0.918} & 3.920 & 3.983 \(\pm\) 0.065 & 3.918 \(\pm\) 0.063 \\
    VARSTok($\tau=0.8$)& 36.81& \textbf{6.787} & \underline{0.899} & \textbf{4.246} & \textbf{4.053 \(\pm\) 0.063} & \textbf{3.946 \(\pm\) 0.062} \\
    VARSTok($\tau=0.7$)& 30.95& \underline{7.294}& 0.895& \underline{4.199} & \underline{4.036 \(\pm\) 0.065} & \underline{3.941 \(\pm\) 0.065} \\
    VARSTok($\tau=0.6$)& 26.29& 9.393 & 0.880 & 4.083 & 3.986 \(\pm\) 0.064 & 3.913 \(\pm\) 0.066 \\
    \bottomrule
    \end{tabular}%
  \label{tab:TTS}%
\end{table*}%

To better understand how VARSTok allocates tokens over time, we visualize the token boundaries for two speech segments in Figure \ref{fig:visualization}.  As illustrated, the 75 Hz and 40 Hz WavTokenizer assigns tokens uniformly across time regardless of the acoustic structure, leading to potential over- or under-tokenization in regions of varying complexity. In contrast, VARSTok dynamically adjusts token duration in response to local feature similarity: shorter tokens are assigned to acoustically rich regions (e.g., rapid formant transitions at 0.15-0.20s in (a)), while longer tokens are assigned to redundant regions (e.g., steady vowels at 0.35-0.40s of (a) or silence at 0.35-0.50s of (b)). This adaptive allocation enables improved token efficiency while preserving critical acoustic details, as reflected in higher UTMOS and PESQ scores under comparable or lower frame rates. These qualitative findings support our design motivation: variable token durations aligns more closely with the temporal dynamics of speech and yields a more semantically faithful and resource-efficient representation.

\subsection{Semantic Evaluation}

We adopted ARCH benchmark to assess the semantic quality of the learned speech representations. We report classification accuracy and macro-averaged F1 score in Table \ref{tab:arch}. Despite operating at a lower average token rate, VARSTok achieves superior performance compared to the 40 Hz WavTokenizer across all tasks and all settings.
These results demonstrate that dynamic token allocation leads to more semantically expressive representations.

\subsection{TTS Language Modeling}



 To evaluate the compatibility of VARSTok with downstream speech LMs, we trained an TTS model based on a decoder-only autoregressive architecture.
Similar to VALL-E \cite{chen2025neural}, given an input text and a 3-second speech prompt from the target speaker, the model autoregressively generates speech tokens, which are then directly decoded into waveform using the the decoder of the pretrained speech tokenizer. 
As shown in Table \ref{tab:TTS}, VARSTok with $\tau=0.8$ and $S_{\max}=4$ achieves the best overall performance despite using fewer tokens on average, improving naturalness while reducing WER compared to the 40 Hz WavTokenizer baseline. 
Even at a lower token rate of 30.95 Hz, our model achieves comparable WER and higher UTMOS, demonstrating that VARSTok enhances both efficiency and generation quality. 
We note a slight decrease in the objective speaker similarity score as the token rate drops. However, such metrics are known to sometimes diverge from human perception.
Subjective results from human evaluators confirm that VARSTok matches the baseline in perceived speaker similarity and naturalness, indicating the minor drop in the objective SIM score is not perceptually significant. These results reinforces the overall superiority of our dynamic tokenization approach for enabling token-efficient and high-fidelity speech synthesis.
Reprensentative speech samples are provided in the Appendix A.

\section{Conclusion}
We introduce VARSTok, a variable-frame-rate speech tokenizer that produces compact token sequences by dynamically segmenting speech using a temporal-aware clustering algorithm. To support seamless integration with downstream speech LMs, we incorporate a implicit duration coding scheme that encodes both content and temporal span into a single token index. 
Comprehensive experiments confirmed that VARSTok enhances token efficiency while maintaining or surpassing the performance of fixed-rate baselines, establishing it as a principled and effective framework for dynamic speech tokenization. Future work could extend this dynamic paradigm to other audio domains such as music.

\section{Acknowledgments}
This work was partially funded  by the National Nature Science Foundation of China under Grant 62301521 and the Anhui Province Major Science and Technology Research Project under Grant S2023Z20004.

\bibliography{aaai2026}

\newpage
\setcounter{secnumdepth}{0} 
\makeatletter
\@ifundefined{isChecklistMainFile}{
  \newif\ifreproStandalone
  \reproStandalonetrue
}{
  \newif\ifreproStandalone
  \reproStandalonefalse
}
\makeatother

\ifreproStandalone
\documentclass[letterpaper]{article}
\usepackage{aaai2026}
\setlength{\pdfpagewidth}{8.5in}
\setlength{\pdfpageheight}{11in}
\usepackage{times}
\usepackage{helvet}
\usepackage{courier}
\usepackage{xcolor}
\frenchspacing

\def\year{2021}\relax
\usepackage{helvet} 
\usepackage{courier}  
\usepackage[hyphens]{url}  
\usepackage{graphicx} 
\urlstyle{rm} 
\def\UrlFont{\rm}  
\usepackage{natbib}  
\usepackage{caption} 
\frenchspacing  
\setlength{\pdfpagewidth}{8.5in}  
\setlength{\pdfpageheight}{11in}  
\usepackage[utf8]{inputenc}
\usepackage{graphicx} 
\usepackage{amsmath,amssymb,gensymb}
\usepackage{booktabs} 

\setcounter{secnumdepth}{0} 

\title{Technical Appendix}

\begin{document}
\maketitle
\else
\clearpage
\twocolumn[
\begin{center}
    {\LARGE \bf Technical Appendix \par}
    \vspace{1em}
\end{center}
]

\addcontentsline{toc}{section}{Technical Appendix} 
\fi

\section*{A. Speech Samples and Implementation Codes}
Speech samples and implementation codes are available at \url{https://anonymous.4open.science/w/VARSTok-demo-8E60/} .

\section*{B. Training Objective of VARSTok}

VARSTok is trained using a multi-term loss function that balances waveform reconstruction fidelity, codebook stability, and perceptual naturalness. We adopt the same loss structure as WavTokenizer, and define the total loss as:
\begin{equation}
\mathcal{L} = \lambda_{mel} \mathcal{L}_{mel} + \lambda_q \mathcal{L}_q + \lambda_{adv} \mathcal{L}_{adv} + \lambda_{feat} \mathcal{L}_{feat}.
\end{equation}
where:

\begin{itemize}
    \item Mel-spectrogram reconstruction loss $\mathcal{L}_{mel}$: We measure the L1 distance between the mel-spectrograms of the original waveform $\mathbf{x}$ and the reconstructed waveform $\hat{\mathbf{x}}$. 
\begin{equation}
    \mathcal{L}_{mel} = ||\text{Mel}(\mathbf{x})-\text{Mel}(\hat{\mathbf{x}})||_1
\end{equation}
This loss enforces time-frequency alignment between real and generated signals, capturing overall speech structure and prosody.
    \item Vector quantization Loss $\mathcal{L}_q$: Instead of optimizing the full VQ-VAE loss which includes codebook loss, we follow WavTokenizer and only optimize the commitment loss, which encourages encoder outputs to remain close to the codebook entries:
\begin{equation}
    \mathcal{L}_q=\sum_{n=1}^N||\mathbf{z}_n-\text{sg}[\mathbf{e}_{k_n}]||,
\end{equation}
where $z_n$ is the embedding of cluster $C_n$, $\mathbf{e}_{k_n}$ is the assigned codebook vector, and $\text{sg}[\cdot]$ denotes the stop-gradient operator. The codebook vector is updated through exponential moving average (EMA) with random awakening to improve utilization and avoid collapse.
    \item Adversatial Loss $\mathcal{L}_{adv}$: To improve perceptual quality, we adopt a hinge-based GAN loss using multiple discriminators $\{D_k\}_{k=1}^{K}$:
\begin{equation}
    \mathcal{L}_{adv} = \frac{1}{K}\sum_{k=1}^K\max(0, 1-D_k(\hat{\mathbf{x}})).
\end{equation}
We inherit the discriminator architecture and training pipeline from WavTokenizer, including the multi-period and multi-resolution STFT-based discriminators. This encourages the decoder to produce waveforms indistinguishable from real speech across multiple time scales and resolutions.
    \item Feature Matching Loss $\mathcal{L}_{feat}$: To stabilize GAN training and preserve structural consistency, we apply a feature matching loss over discriminator feature maps:
\begin{equation}
    \mathcal{L}_{feat} = \frac{1}{K\cdot L}\sum_{k=1}^K \sum_{l=1}^L||D_k^{(l)}(\mathbf{x})-D_k^{(l)}(\hat{\mathbf{x}})||_1,
\end{equation}
\end{itemize}
where $D_k^{(l)}(\hat{\mathbf{x}})$ denotes the $l$-th intermediate layer output of discriminator $D_k$. This penalizes discrepancies in internal discriminator representations between real and synthesized speech.

The loss weights are set as $\lambda_{mel} = 45$, $\lambda_q = 1$, $\lambda_{adv} = 1$, and $\lambda_{feat} = 4$, consistent with the original WavTokenizer configuration, ensuring stable training while balancing fidelity and perceptual quality.

\section*{C. Training Configuration Details}
\label{sec:appendix_training_details}

All our models were trained on the 585-hour LibriTTS corpus using a total batch size of 128, distributed across 4 NVIDIA A800 GPUs. Each sample in the batch consisted of a 3-second audio chunk. For optimization, we employed the Adam optimizer with default hyperparameters ($\beta_1=0.9, \beta_2=0.999$) and applied gradient clipping with a maximum L2 norm of 1.0 to ensure stability. The models were trained for 60 epochs under a two-stage learning rate schedule: an initial rate of 1e-4 for the first 30 epochs, followed by a reduction to 2e-5 for the next 30 epochs. This staged annealing strategy facilitated effective convergence by allowing for rapid initial learning followed by fine-tuning.

\section*{D. Baseline Model Details}
This section provides detailed configurations for the additional baseline models used in our speech reconstruction evaluation, namely BigCodec, XCodec2.0, DAC, and SQCodec.

\subsection{D.1 BigCodec}
We trained the BigCodec from scratch on the 24 kHz LibriTTS corpus using its official codebase \footnote{\url{https://github.com/Aria-K-Alethia/BigCodec}.}. To achieve the target 40 Hz frame rate, the encoder and decoder strides were configured as [6, 5, 5, 4].

\subsection{D.2 XCodec2.0}
The XCodec2.0 was also trained from scratch on the 24 kHz LibriTTS corpus using its official codebase \footnote{\url{https://github.com/zhenye234/X-Codec-2.0/}}. We adapted its official implementation to support 24 kHz speech by adjusting the encoder stride to [2, 3, 4, 4, 5] and modifying the decoder accordingly. It should be noted that XCodec2.0 is a hybrid tokenizer that incorporates semantic features from a w2v-BERT-2.0 model during both training and inference.

\subsection{D.3 SQCodec}
We evaluated SQCodec using its publicly released checkpoint \footnote{\url{https://github.com/zhai-lw/SQCodec}.}, as the official training code was not available. This model operates on 16 kHz waveforms. Its results are presented as an optimistic reference, as it benefits from two significant advantages: its checkpoint was trained on datasets substantially larger than LibriTTS, and it performs compression on 16kHz audio, which is an inherently simpler task compared to the 24kHz standard used by the other models undern similar bitrates.

\section*{E. ARCH Benchmark (Speech Portion)}
To evaluate the semantic quality of learned speech representations from different tokenizers, we adopt the speech portion of the ARCH benchmark, which offers a standardized evaluation pipeline across multiple speech classification tasks.

\subsection{E.1 Dataset}
We utilize the four speech-forcused datasets for evaluation:
\begin{itemize}
    \item EMOVO: Emotion classification in Italian with 7 classes.
    \item RAVEDSS: Emotion classification with 8 classes.
    \item  AudioMNIST: Spoken digit recognition with 10 classes.
    \item SLURP: Intent classification with 77 classes.
\end{itemize}
All are single-label classification tasks, with average sample durations ranging from 0.6 to 3.7 seconds.

 \subsection{E.2 Evaluation Protocol}
 We follow ARCH’s frozen representation evaluation protocol. Specifically, we first extract frame-level token embeddings from our variable-frame-rate tokenizer and apply average pooling over time to obtain a fixed-dimensional embedding for each audio sample. A linear classifier is then trained on top of these embeddings for 200 epochs using the AdamW optimizer with a learning rate of 0.001, employing a linear warmup schedule followed by linear decay. Throughout this process, the tokenizer itself remains frozen without any fine-tuning, ensuring that the evaluation reflects the intrinsic representational quality of the learned tokens. Consistent with ARCH practices, we report classification accuracy and macro-averaged F1 scores on the test splits for each dataset.

 \section*{F. Decoder-Only TTS Model Training Configuration}
 This section provides the implementation and training details for the autoregressive TTS model used for our downstream TTS evaluation. Our model follows a decoder-only architecture, similar to the autoregressive component of VALL-E, but does not incorporate a non-autoregressive decoder for predicting residual VQ tokens. The implementation is adapted from a publicly available project \footnote{\url{https://github.com/lifeiteng/vall-e}.}. 

\subsection{F.1 Model Architecture}
The TTS model is built on a Transformer decoder consisting of 12 layers, 16 attention heads, and an embedding dimension of 1024. The model is conditioned on a sequence of phonemes derived from the input text and a 3-second speech prompt, and it autoregressively predicts the output speech token sequence. Unlike the full VALL-E architecture, our model does not include multi-stage residual token prediction or non-autoregressive refinement modules.

\subsection{F.2 Training Configuration}
All training was conducted on the 585-hour LibriTTS dataset, with speech uniformly sampled at 24kHz. The model was trained for 100 epochs on a single NVIDIA A800 GPU. To optimize memory usage and training efficiency, we employed a dynamic batching strategy where speech segments of similar length were grouped into the same batch. The model was trained using a standard cross-entropy loss, optimized with the Adam optimizer. We initialized the training with a base learning rate of 0.05 and applied a cosine decay schedule for gradual adjustment as training progressed.

\subsection{F.3 Inference}
During inference, the model autoregressively generates a sequence of speech tokens based on the input phonemes and the 3-second acoustic prompt. The generated token sequence is then passed to the speech tokenizer's decoder to synthesize the final waveform. Model performance was evaluated on a held-out test set, with a focus on assessing the naturalness and intelligibility of the synthesized speech.

\begin{table*}[tbp]
  \centering
  \caption{Speech reconstruction performance on the LibriTTS test-other subset. The bitrate is calculated as $\text{Frame Rate}\times\log_2(K\cdot S_{\max})$ for VARSTok and $\text{Frame Rate}\times\log_2(K)$ for baselines. 
  The optimal results below 40Hz frame rate are marked in \textbf{bold}.  }
    \begin{tabular}{c|c|c|cccc}
    \toprule
    Model & Frame Rate/Hz$\downarrow$ & Bitrate/kbps$\downarrow$ & UTMOS$\uparrow$ & PESQ$\uparrow$  & STOI$\uparrow$  & V/UV F1$\uparrow$ \\
    \hline
    GT    & /& /& 3.4831& /& /& /\\ \hline
    DAC   & 100.00& 1.00& 1.4986& 1.2454& 0.7505& 0.7775\\
    WavTokenizer & 75.00& 0.90& 3.4312& 2.2614& 0.8907& 0.9172\\
    Xcodec2.0 & 50.00& 0.80& 3.1489& 1.8250& 0.8606& 0.8969\\ 
    VARSTok($\tau=0.7$,$S_{\max}=2$)& 46.50& 0.60& 3.5246& 1.9209& 0.8620& 0.9052\\
    SQCodec & 44.44 & 0.75 & 3.4985  & 1.8120 & /& 0.8967\\  \hline
    BigCodec & 40.00& 0.52 & 3.4125& 1.7219& 0.8462& 0.8968\\
    WavTokenizer & 40.00& 0.48 & 3.0544& 1.6625& 0.8337& 0.8952\\
    VARSTok($\tau=0.8$,$S_{\max}=4$)& 36.57& 0.52 & \textbf{3.5025}& \textbf{1.7544}& \textbf{0.8475}& \textbf{0.8987}\\
    VARSTok($\tau=0.7$,$S_{\max}=4$)& 30.69& 0.43 & 3.4149& 1.6105& 0.8251& 0.8888\\
    VARSTok($\tau=0.6$,$S_{\max}=4$)& 25.97& 0.36& 3.3584& 1.4857& 0.8018& 0.8806\\
    VARSTok($\tau=0.7$,$S_{\max}=8$)& 21.81& 0.33& 3.1834& 1.3875& 0.7791& 0.8663\\
    
    \bottomrule
    \end{tabular}%
  \label{tab:reconstruction-test-other}%
\end{table*}%

\begin{figure*}[t]
    \centering
    \includegraphics[width=\linewidth]{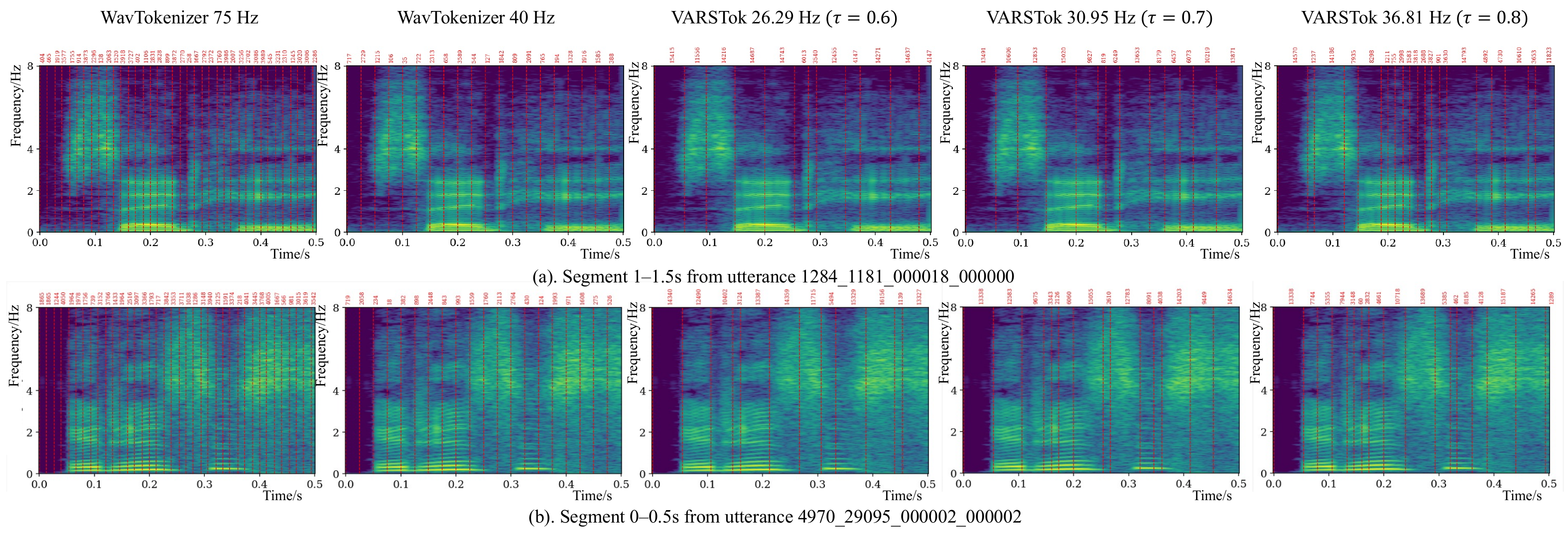}
    \caption{
        Token boundary visualization on two speech segments for wavtokenizer and VARSTok. Vertical red lines indicate token boundaries and token IDs are annotated above. 
    }
    \label{fig:visualization_app}
\end{figure*}

\section*{G. Speech Reconstruction Evaluation on LibriTTS Test-Other Subset}
To further assess the robustness of our method under challenging acoustic conditions, we evaluate VARSTok on the LibriTTS test-other subset, which contains noisier and more diverse recordings than test-clean. As shown in Table \ref{tab:reconstruction-test-other}, VARSTok consistently outperforms or matches strong fixed-rate baselines across all metrics at frame rate below 40 Hz, while operating at a significantly lower bitrate. 
These results highlight VARSTok’s ability to maintain naturalness and clarity even under degraded conditions. Compared to fixed-rate baselines such as WavTokenizer and BigCodec which operate at similar or higher bitrates, VARSTok provides a more efficient and robust representation of speech.

\section*{H. Supplementary Visualizations of Tokenization Results}

Figure~\ref{fig:visualization_app} provides additional comparisons for two new speech samples, further analyzing the influence of the threshold $\tau$ on tokenization results. VARSTok aligns more closely with the temporal fluctuations of the speech signal compared to WavTokenizer, which uses a fixed token rate regardless of speech content. By adapting the token allocation based on local feature similarity, VARSTok captures the underlying dynamics of the speech signal more effectively, especially in regions with rapid transitions or expressive intonation.

The adjustment of $\tau$ introduces a trade-off between token efficiency and granularity. As $\tau$ increases, the clustering process becomes stricter, resulting in more tokens being allocated to the speech signal. This allows VARSTok to capture finer acoustic details, such as the 0.2-0.25s segment in (a) and the 0.15-0.2s segment in (b). However, this comes at the cost of a higher frame rate, as more tokens are required to represent the same speech segment. Therefore, higher $\tau$ values increase the number of tokens and the computational cost, while potentially providing a more detailed representation of the speech signal.

This trade-off between the number of tokens and the frame rate allows VARSTok to adjust its tokenization strategy to balance efficiency and accuracy. A lower $\tau$ may lead to fewer tokens and bette efficiency, but at the expense of finer details in the speech representation. Conversely, increasing $\tau$ enhances tokenization detail but also increases the computational burden. These visualizations highlight the flexibility of VARSTok, showing how it can dynamically adjust to different speech characteristics while maintaining an effective balance between token efficiency and representational quality.

\begin{table*}[h!]
\centering
\caption{Subjective evaluation of TTS performance. We report Mean Opinion Score (MOS) for naturalness and Similarity MOS (SMOS) for speaker similarity, with 95\% confidence intervals. The optimal results are marked in \textbf{bold}.  }
\label{tab:tts_subjective}
\begin{tabular}{c|c|ccc}
\toprule
Tokenizer & Frame Rate (Hz) & MOS \(\uparrow\) & SMOS \(\uparrow\) & RTF \(\downarrow\) \\
\hline
WavTokenizer & 40.00 & 3.983 \(\pm\) 0.065 & 3.918 \(\pm\) 0.063 & 0.766 \\
VARSTok ($\tau=0.8$) & 36.81 & \textbf{4.053 \(\pm\) 0.063} & \textbf{3.946 \(\pm\) 0.062} & 0.716 \\
VARSTok ($\tau=0.7$) & 30.95 & 4.036 \(\pm\) 0.065 & 3.941 \(\pm\) 0.065 & 0.602 \\
VARSTok ($\tau=0.6$) & 26.29 & 3.986 \(\pm\) 0.064 & 3.913 \(\pm\) 0.066 & \textbf{0.487} \\
\bottomrule
\end{tabular}
\end{table*}

\begin{table*}[h!]
\centering
\caption{Analysis of VARSTok with different VQ codebook sizes on LibriTTS test-clean. ``Exp. Space" and ``Exp. Usage" refer to the expanded token space and its utilization, respectively. All models are trained with \(\tau=0.7\) and \(S_{\max}=4\). Codebook Usage refers to the percentage of utilized entries in the original VQ codebook. Expanded Token Space Usage refers to the percentage of utilized entries in the conceptual vocabulary of size \(K \cdot S_{\max}\).}
\label{tab:codebook_analysis}
  \setlength{\tabcolsep}{1mm} 
\begin{tabular}{cc|cc|cc|cccc}
\toprule
$K$& Exp. Space & Codebook Usage & Exp. Usage & Frame Rate/Hz & Bitrate/kbps & UTMOS $\uparrow$ & PESQ $\uparrow$ & STOI $\uparrow$ & V/UV F1$\uparrow$ \\
\hline
2048 & 8192 & 100\%& 99.85\%& 30.96& 0.40& 3.8262& 1.6403& 0.8519& 0.8992\\
4096 & 16384 & 100\% & 99.16\% & 30.95  & 0.43  & 3.8949 & 1.7095 & 0.8601 & 0.9047 \\
8192 & 32768 & 72.72\% & 72.19\% & 30.94  & 0.46  & 3.9042 & 1.7169 & 0.8625 & 0.9054 \\
16384 & 65536 & 40.17\% & 39.40\% & 31.16  & 0.50  & 3.9390 & 1.7559 & 0.8647 & 0.9065 \\
\bottomrule
\end{tabular}%
\end{table*}

\section*{I. Subjective Evaluation of TTS Performance}

To complement the objective metrics, we conducted a subjective listening test to evaluate the naturalness and speaker similarity of the synthesized speech in the zero-shot TTS evaluation task. We randomly selected 25 samples from the TTS test set and synthesized them using the trained TTS model conditioned on different tokenizers. The evaluation was carried out on the Amazon Mechanical Turk\footnote{\url{https://www.mturk.com/}.} platform with 35 native English speakers. For naturalness, listeners rated the speech on a 5-point Mean Opinion Score (MOS) scale. For speaker similarity, they rated how similar the synthesized voice was to the original speech prompt on a 5-point Similarity MOS (SMOS) scale. 

The results, presented in Table~\ref{tab:tts_subjective}, strongly support our findings from the objective evaluations and reveal important insights. All variants of VARSTok achieve MOS and SMOS scores that are competitive with or superior to the 40 Hz WavTokenizer baseline, despite operating at significantly lower average frame rates.

Notably, the results reveal a clear optimization trade-off with respect to the clustering threshold $\tau$.  The model configured with \(\tau=0.8\) (36.81 Hz) achieves the highest scores in both naturalness (MOS=4.053) and speaker similarity (SMOS=3.946), outperforming the fixed-rate baseline. As \(\tau\) decreases, leading to more aggressive compression and lower frame rates, the perceptual quality slightly declines but remains on par with the baseline even at just 26.29 Hz. This suggests that a moderately conservative clustering threshold like \(\tau=0.8\) best preserves the crucial acoustic cues for naturalness and speaker identity. These results provide compelling evidence that VARSTok not only improves token efficiency but also enhances the perceptual quality of synthesized speech, with the choice of \(\tau\) offering a clear trade-off between compression rate and fidelity.

\section*{J. Analysis of Downstream TTS Inference Efficiency}
\label{sec:appendix_efficiency}

A potential concern regarding VARSTok is that its implicit duration-content encoding scheme expands the token vocabulary size for the downstream TTS language model from $K$ to $K \times S_{\max}$. This could increase the computational cost per decoding step due to a larger final projection layer. To investigate this trade-off, we measured the Real-Time Factor (RTF) of our TTS model. RTF is defined as the processing time required to synthesize one second of speech; a lower value indicates higher efficiency.

The results, presented in the last column of Table~\ref{tab:tts_subjective} clearly show that the substantial efficiency gains from processing significantly shorter token sequences (a consequence of a lower average frame rate) far outweigh the computational drawback of an expanded vocabulary. Specifically, all VARSTok variants achieve a lower RTF compared to the 40 Hz WavTokenizer baseline. This trend becomes more pronounced with increased compression: the model configured with $\tau=0.6$ achieves an RTF of 0.487, representing a 36\% inference speedup over the baseline, while maintaining comparable perceptual quality. These findings provide strong evidence that VARSTok not only improves compression and perceptual quality but also robustly enhances the end-to-end computational efficiency of downstream generative tasks.

\section*{K. Analysis of Codebook Size and Utilization}
To investigate the impact of the VQ codebook size $K$ on the rate-quality trade-off and model efficiency, we conducted an ablation study with different values of $K$, while keeping other hyperparameters fixed (\(\tau=0.7, S_{\max}=4\)). The results are summarized in Table~\ref{tab:codebook_analysis}.

The results reveal a clear trade-off. As the codebook size $K$ increases from 2048 to 16384, we observe a consistent improvement across all reconstruction metrics. For instance, the UTMOS score rises from 3.8262 to 3.9390, which is expected as a larger codebook provides greater representational capacity to capture fine-grained acoustic details. However, this gain in quality comes at the cost of a higher bitrate increasing from 0.43 kbps to 0.50 kbps and, more importantly, a sharp decline in codebook efficiency.

The codebook utilization, which measures the percentage of used codebook entries, drops dramatically from 100\% at $K=2048$ to a mere 40.17\% at $K=16384$. This indicates that larger codebooks suffer from significant codebook collapse or redundancy, where a majority of entries remain unused. A similar trend is observed in the expanded token space utilization. Notably, the average frame rate remains stable at approximately 31 Hz across different codebook sizes, confirming that the token rate is primarily controlled by our adaptive clustering algorithm rather than the VQ codebook size.

These findings validate our choice of $K=4096$ for the main experiments reported in the paper. This setting strikes an excellent balance between high reconstruction quality, efficient codebook utilization, and a moderate bitrate, avoiding the diminishing returns of larger, under-utilized codebooks.


\end{document}

\end{document}